# Designing a Deep Learning-Driven Resource-Efficient Diagnostic System for Metastatic Breast Cancer: Reducing Long Delays of Clinical Diagnosis and Improving Patient Survival in Developing Countries

William Gao [1], Dayong Wang [2 *], Yi Huang [1]

1 Department of Mathematics and Statistics, University of Maryland, Baltimore County, Baltimore, MD, USA. 2 Meta Platforms Inc., Boston, MA, USA

* Corresponding author: Dayong Wang, Meta Platforms Inc. Email: dayong.wangts@gmail.com

**Abstract**

Breast cancer is one of the leading causes of cancer mortality. Breast cancer patients in developing countries, especially sub-Saharan Africa, South Asia, and South America, suffer from the highest mortality rate in the world. One crucial factor contributing to the global disparity in mortality rate is long delay of diagnosis due to a severe shortage of trained pathologists, which consequently has led to a large proportion of late-stage presentation at diagnosis. The delay between the initial development of symptoms and the receipt of a diagnosis could stretch upwards 15 months. To tackle this critical healthcare disparity, this research has developed a deep learning-based diagnosis system for metastatic breast cancer that can achieve high diagnostic accuracy as well as computational efficiency and mobile readiness suitable for an under-resourced environment. We evaluated four Convolutional Neural Network (CNN) architectures: MobileNetV2, VGG16, ResNet50 and ResNet101. The MobileNetV2-based diagnostic model outperformed the more complex VGG16, ResNet50 and ResNet101 models in diagnostic accuracy, model generalization, and model training efficiency. The visual comparisons between the model prediction and ground truth have demonstrated that the MobileNetV2 diagnostic models can identify very small cancerous nodes embedded in a large area of normal cells which is challenging for manual image analysis. Equally Important, the light weighted MobleNetV2 models were computationally efficient and ready for mobile devices or devices of low computational power. These advances empower the development of a resource-efficient and high performing AI-based metastatic breast cancer diagnostic system that can adapt to under-resourced healthcare facilities in developing countries. This research provides an innovative technological solution to address the long delays in metastatic breast cancer diagnosis and the consequent disparity in patient survival outcome in developing countries.

**Introduction**

Breast cancer is the most diagnosed cancer among women and the fifth leading cause of cancer mortality, accounting for over 2.3 million new cases and 6.9% of all cancer deaths [1]. Breast cancer patients in developing countries, especially sub-Saharan Africa, South Asia, and South America, suffer from the highest mortality rates in the world. One crucial factor contributing to the global disparity in mortality rate is a long diagnosis delay. Research on Turn Around Time (TAT) – the time from biopsy till receipt of report by clinician - from Malawi showed that the median TAT for unpaid samples was 71 days. At two facilities in Rwanda, the delay between the initial development of symptoms and the ultimate receipt of a diagnosis could reach 15 months [2]. Consequently, the long diagnosis delay has led to a large proportion of late-stage presentation at diagnosis. An analysis of 83 studies across 17 sub-Saharan African countries reported that 77% of all staged cases were stage III/IV at diagnosis [3]. The long diagnosis delay is primarily attributed to a severe shortage of trained pathologists. The average number of pathologists per head of population is 1 to 1,000,000 in sub-Saharan regions, compared with the ratio of 1 pathologist to 15,000–20,000 in the US and UK [4]. Another challenge is access to healthcare facilities. In sub-Saharan Africa, more than 170 million people are more than 2 hours from the nearest regional hospital, and about 40% of the population is more than 4 hours away from a national hospital [5]. Additionally, diagnosis quality may be impacted by a lack of resources and training. A retrospective assessment of reports of breast carcinoma in the university teaching hospital in Lagos, Nigeria showed a 46.9% discordance rate in the basic diagnosis when the cases were examined by pathologists in the UK [2].

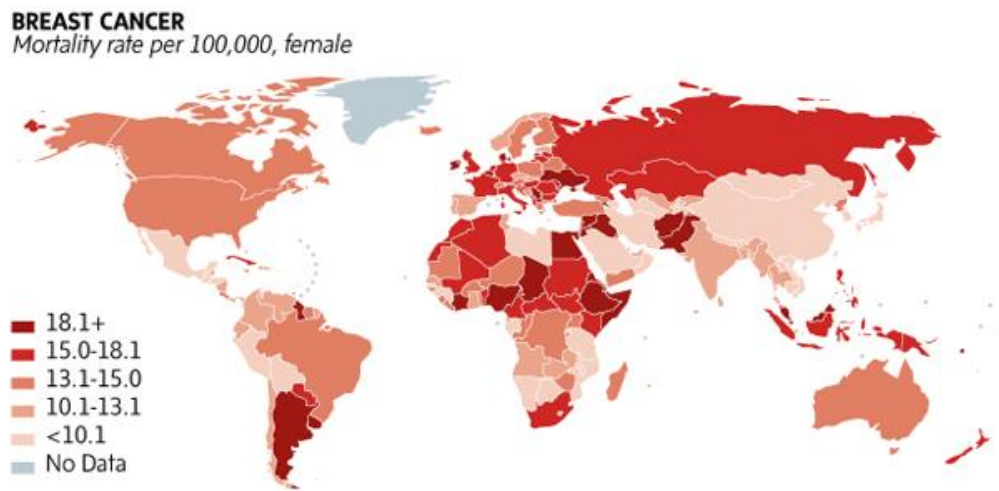

**Figure 1.** World map of breast cancer mortality rates (all ages), showing high mortality rates in sub-Saharan Africa, South Asia, and South America [1]

Artificial intelligence (AI) presents a new opportunity for advancing medicine and healthcare [6]. Currently, pathologists manually review histopathology slides to examine whether cancerous regions are present in tissue samples. However, manual review is time-consuming and subject to human error, especially in healthcare facilities in resource-constrained regions [2]. AI has a great potential to bring changes to cancer diagnosis by providing faster, more accurate, and more robust technological solutions. Machine learning-driven image analysis can assist pathologists by narrowing their search for high-risk areas on the whole slide of a biopsy sample and making their diagnosis more accurate and efficient. Progress has been made to develop machine learning-driven image analysis methods for breast cancer diagnosis [7, 8, 9]. However, real-world clinical application of these methods, especially in remote and under-resourced areas, remains a major challenge. The complex neural network architecture, computational cost, and lack of mobile capacity make the existing methods less feasible and adaptive to a geographically distributed and under-resourced environment. The objective of this research is to develop a deep learning-driven resource-efficient diagnostic system for metastatic breast cancer that can provide a standardized and automated platform to help pathologists to expedite diagnosis. To address the challenges encountered by the healthcare facilities in the remote and resource-constrained regions, our objective is to enable the system to pursue computational efficiency and mobile readiness while achieving high diagnostic accuracy.

## 2. Methods
## 2.1. Framework of Deep Learning-Driven Diagnostic System for Metastatic Breast Cancer

The deep learning-driven diagnostic system for metastatic breast cancer consists of three major components: histopathological image pre-processing, training to build deep learning-driven diagnostic models, and testing for prediction on unseen whole slide images (WSIs) (Figure 2).

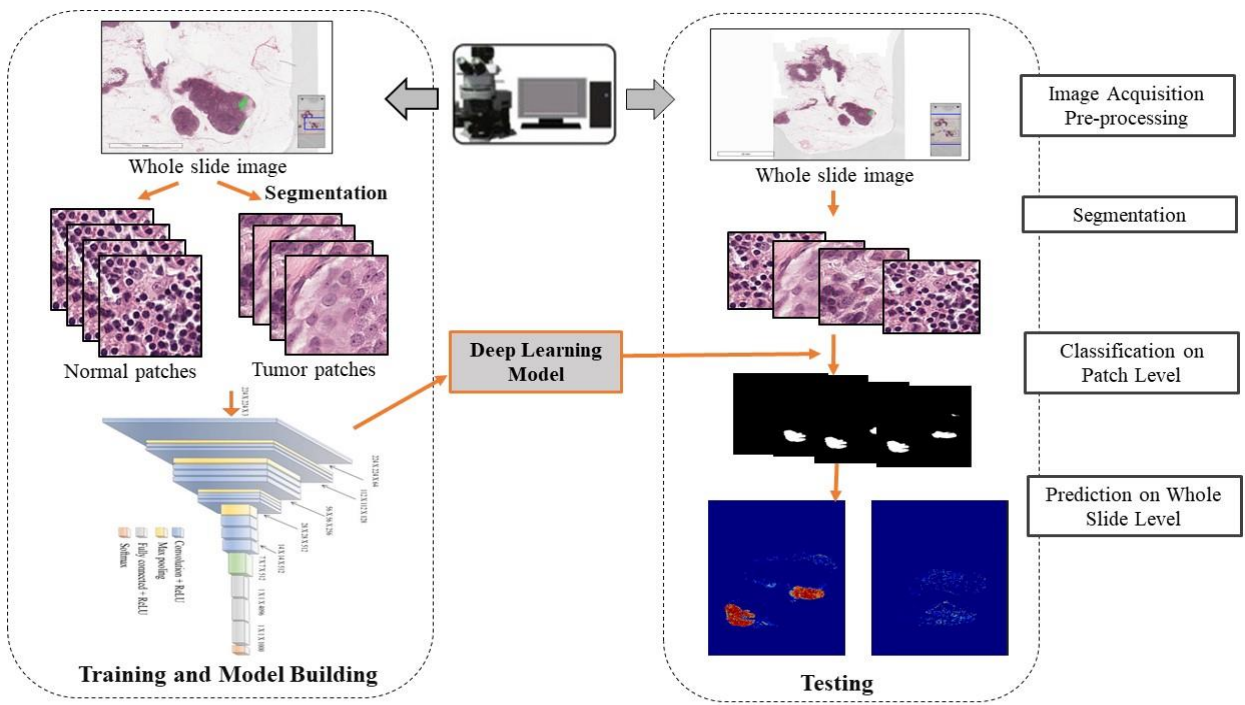

**Figure 2.** The workflow of deep learning-driven diagnostic system for metastatic breast cancer

### 2.1.1. Histopathological Image Pre-Processing

The diagnostic models were trained based on the ground truth data, i.e., whole slide images (WSI) of sentinel lymph nodes with a pathologist's delineation of regions of metastatic cancer. The image normalization was conducted with the WSI Color Standardization procedure [10] to minimize potential variations in the color and intensity of Hematoxylin and Eosin (H&E) staining. Tissue areas within the normalized WSIs were identified and extracted using a threshold-based segmentation method [11]. The mask images were generated for model training from the HSV representation transformed from the original RGB images [12].

### 2.1.2. Patch-Based Diagnostic Model Building

Model training used WSIs and the ground truth image annotation indicating the delineation of cancerous regions as input data. The WSIs were randomly extracted into a large number of small patches. Using the small patches, the deep learning model was trained to recognize cancerous cells on a small scale, improving data volume and model training efficiency. These patches extracted from each WSI were categorized as a positive tumor, negative tumor, and negative normal. A positive tumor patch was extracted from a tumor slide, containing cancerous regions; a negative tumor patch was extracted from a tumor slide but did not contain a cancerous region; a negative normal patch was extracted from a noncancerous/normal slide. During patch extraction, masks were created using a lower and upper bound of pixel color. After reducing mask noise with the "opening" and "closing" morphology processes, the pixel values of each mask were used to categorize the patch.

Four Convolutional Neural Network architectures, i.e., MobileNetV2, VGG16, ResNet50 and ResNet101, were used for building diagnostic models. The four CNN architectures were assessed for diagnosis performance and computational efficiency. The basic characteristics of the CNN architectures are summarized in section 2.2. The diagnostic models were trained to discriminate between cancerous and noncancerous patches using a large number of small positive and negative patches randomly extracted from the set of training WSIs. Five-fold cross-validation and independent (unseen) image data were used to evaluate diagnostic performance.

### 2.1.3. Whole Slide-Based Diagnosis

As shown in Figure 2, unseen WSIs were pre-processed and segmented following the procedures described in sections 2.1.1. The diagnostic models were used to discriminate the cancerous vs. non-cancerous patches. With the patch-level classification, a tumor probability heatmap was generated for each WSI to indicate the probability of each pixel being cancerous, with each pixel containing a value between 0 and 1, indicating the probability that the pixel contains tumor.

### 2.1.4. Diagnostic Performance Evaluation

Diagnostic performance was evaluated using accuracy and receiver operating characteristics (ROC) curve. Accuracy is defined as the number of correct predictions divided by total number of predictions. ROC curve plots true positive rate (sensitivity) vs. false positive rate (1-specificity); Area under ROC curve (ROC AUC) provides an aggregate measure of performance across all possible classification thresholds.

## 2.2. Convolutional Neural Network Architectures

Four deep learning network architectures were evaluated for diagnostic performance for metastatic breast cancer and computational efficiency, including MobileNetV2, VGG16, ResNet50 and ResNet101. MobileNetV2 is a CNN architecture that aims to perform computer vision tasks efficiently on mobile devices. It is designed to achieve efficiency by incorporating the Inverted Residual Structure and the Depthwise Separable Convolution to significantly reduce the model size and complexity [13]. VGG16 is a CNN architecture that is designed for large-scale image recognition. VGG16 improves accuracy by increasing the depth to 16 weight layers using an architecture with very small (3×3) convolution filters, in combination with more non-linear activation layers for better discriminative decision function [14]. ResNet is a Residual Network, a category of CNN architectures that is intended to improve model accuracy by stacking layers to enrich the features of the model [15]. Resnet50 is a variant of ResNet whose neural network layers reach a depth of 50, with a bottleneck design to reduce the time taken to train the layers. ResNet101 is a large Residual Network with 101 neural network layers. In Keras Applications, MobileNetV2 uses 3.5 million (M) of parameters and has a size of 14MB; VGG16 uses 138.4M parameters and has a size of 528MB; ResNet50 uses 25.6M parameters and has a size of 98MB; ResNet101 uses 44.7M parameters and has a size of 171MB [16].

## 2.3. Datasets

The data consists of 222 whole slide histopathological images from the 2016 Camelyon ISBM challenge [9]. The slides contain sentinel lymph node tissues extracted by the Radboud University Medical Center (Nijmegen, the Netherlands), as well as the University Medical Center Utrecht (Utrecht, the Netherlands).

## 3. Results

### 3.1. Evaluation of Deep Learning Network Architectures: Diagnostic Accuracy and Computational Efficiency

To identify a CNN architecture that is suitable for a resource-constrained environment, four CNN architectures were evaluated for diagnostic performance and computational cost, including MobileNetV2, VGG16, ResNet50 and ResNet101. The diagnostic models were trained and assessed for accuracy using five-fold cross-validation. For an individual fold, each model ran through ten iterations or "epochs". Each epoch allowed the model to reevaluate its weights to determine a more effective set of values. With a stochastic gradient descent (SGD) optimizer, the weights were retrained or optimized for the data with a 0.01 learning curve. Model accuracy was assessed for the training set and validation sets at each epoch. For training, the accuracy of MobileNetV2 (mean ± standard deviation: 0.963 ± 0.005) was higher than VGG16 (0.940 ± 0.006), ResNet50 (0.847 ± 0.015) and ResNet101 (0.851 ± 0.015). Similarly, the cross-validation results showed that the accuracy of MobileNetV2 (0.885 ± 0.043) was similar with VGG16 (0.882 ± 0.058) but higher than ResNet50 (0.701 ± 0.072) and ResNet101 (0.707 ± 0.076).

The independent testing of the four diagnostic models with different CNN architectures was conducted using an independent testing set. The accuracies of MobileNetV2 (0.903) and VGG16 (0.898) were substantially higher than ResNet50 (0.721) and ResNet101 (0.857). Receiver operating characteristic (ROC) analysis was performed and ROC Area Under curve (AUC) was used to measure the diagnostic performance. As shown in Figure 3A, the ROC AUC of MobilenetV2 (0.933, 95% confidence interval (CI): 0.930 - 0.936) was higher than VGG16 (0.911, 95% CI: 0.908- 0.915), Resnet50 (0.869, 95% CI: 0.866 - 0.873), and Resnet101 (0.873, 95% CI: 0.869 - 0.876).

The models were also evaluated for time per inference step, a measure of computational efficiency (Figure 3B). The time per inference step for the MobileNetV2 model was 15ms/step, which was substantially lower than that of VGG16 (48ms/step), ResNet50 (37ms/step), and ResNet110 (56ms/step). The result suggests that the MobileNetV2 model was more computationally efficient than the models with more complex CNN architectures.

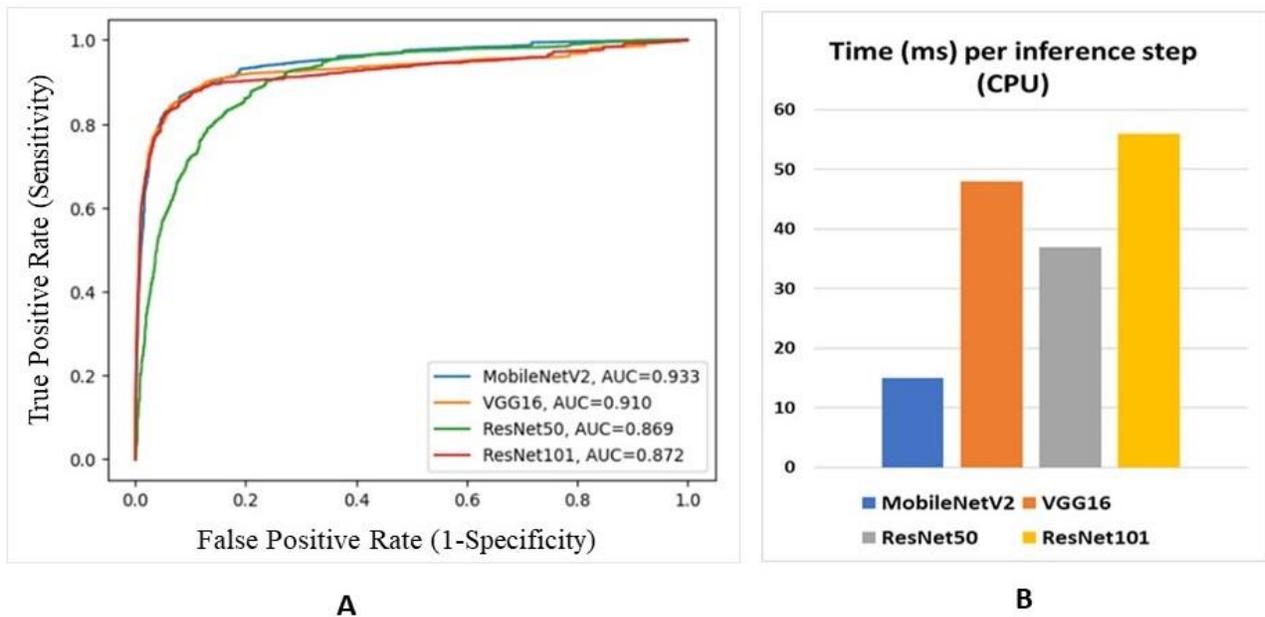

Figure 3. (A) Receiver Operating Characteristics (ROC) curves for independent set testing: comparison of ROC AUC among the CNN architectures MobileNetV2, VGG16, ResNet50 and ResNet101. (B) Comparison of time per inference step among the four CNN architectures.

### 3.2. Data Augmentation: Assessment of Model Generalization

Model generalization capacity is important for a model to properly adapt to previously unseen data and to achieve reasonable prediction performance. It is noticeable that the accuracy values of both the MobileNetV2-based and VGG16-based models exceeded 0.880 in the cross-validation while those of ResNet50 and ResNet101 dropped substantially in the cross-validation as compared with training. The result implies that the MobileNetV2-based and VGG16-based models had a stronger ability of model generalization as compared with ResNet50 and ResNet101. We used data augmentation techniques to further evaluate model generalization of these CNN architectures. Data augmentation involves training models on data that has undergone various transformations [17]. This experiment was performed to evaluate whether the models can properly generalize to new instances. The image data transformation for data augmentation included 20-degree rotation, 20% zoom, horizontal flip, and vertical flip (Figure 4A).

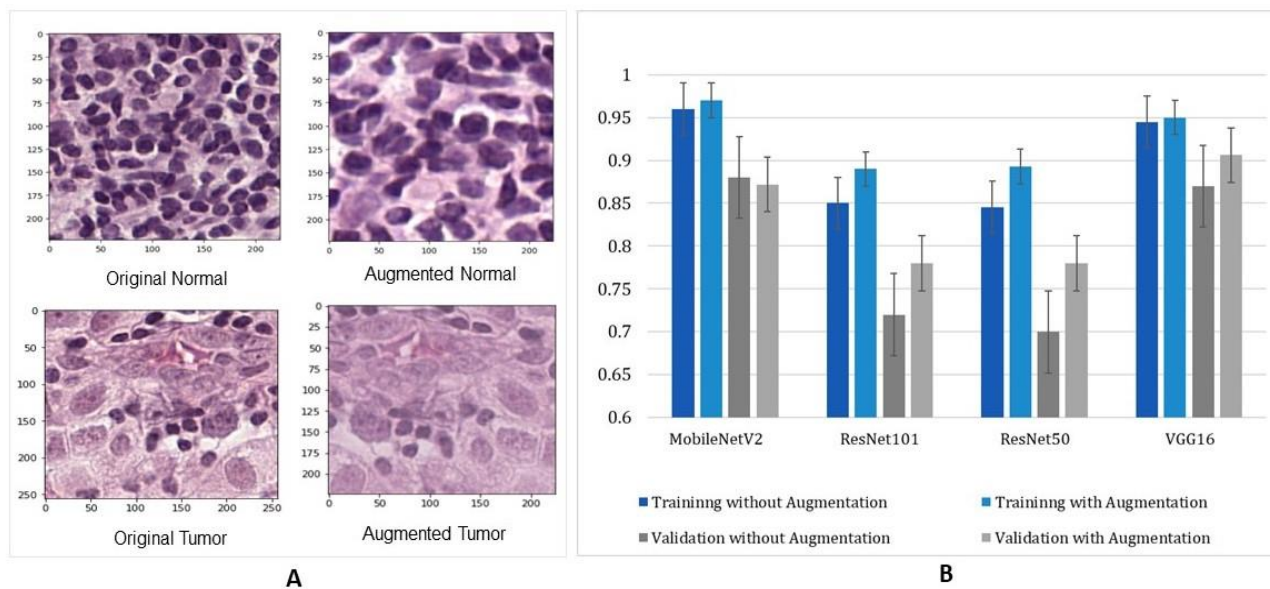

**Figure 4.** A: Image transformation for data augmentation; B: Effects of data augmentation on the accuracy of the classification models with different network structures.

As shown in Figure 4B, with or without data augmentation, the accuracies of MobileNetV2 and VGG16 models were substantially higher than those of ResNet50 and ResNet101 on both training and validation. Additionally, the models based on MobileNetV2 and VGG16 showed a substantially smaller gap in accuracy between training and validation as compared with other models, indicating a stronger generalization ability of the MobileNetV2 and VGG16 models. Data augmentation only caused slight changes to the accuracies of the MobileNetV2 and VGG16 models, while it drastically increased the accuracies of ResNet50 & ResNet101. These results suggest that the MobileNetV2 and VGG16 models were less reliant on bigger data volumes for achieving their peak performance.

### 3.3. Predictive Diagnosis on Unseen Whole Slide Images

The MobileNetV2-based diagnostic model classified small image patches to cancerous vs. non-cancerous with high accuracy. By iterating the model over each patch in the whole slide image, a probability rating was generated for each patch. After completion of the patch-based classification stage, the patch-level predictions were aggregated to create tumor probability heatmap for whole slide-based prediction. The likelihood of cancer presence on a whole slide image was shown on the heatmap, where a pixel represents a patch, and a value represents the probability of the patch containing cancer with the range of 0 ~ 1. The threshold of 0.9 was used to control potential false positives.

Visual comparisons between the MobileNetV2 model prediction and the pathologist's diagnosis (ground truth) were shown in Figure 5. For tumor case A, the cancerous regions predicted by the model were consistent with those delineated by pathologists' diagnosis. Tumor case B was a more complicated case because the cancerous regions were small and embedded in a large area of normal cells. The MobileNetV2 model successfully identified the small cancerous node, and the prediction result was consistent with the pathologists' diagnosis. Identifying such a small cancerous region is challenging in the manual diagnosis process. The results demonstrated that the MobileNetV2-based diagnostic model's capacity in identifying the small regions of high risk and preventing false negative diagnoses.

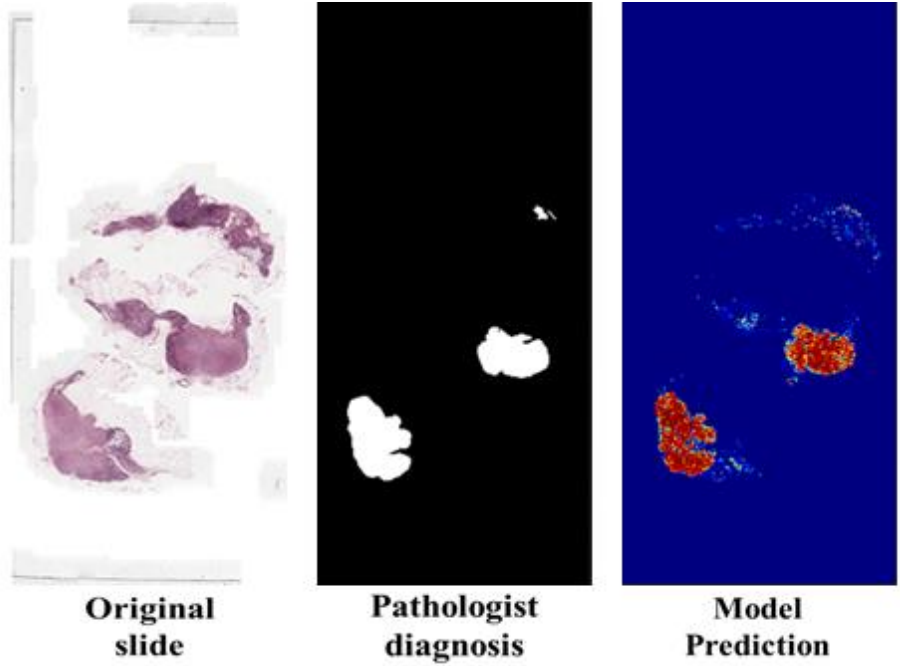
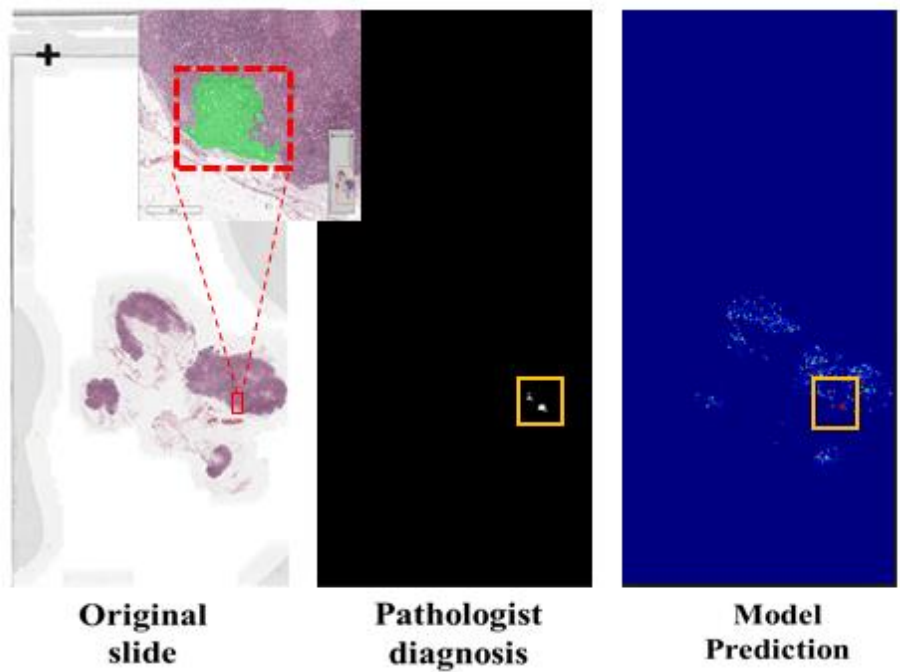

Figure 5. Visual comparisons between the pathologist's diagnosis (ground truth) and the model prediction on cancerous regions on Whole Slide Images.

## 4. Discussion

The recent research and development for computer-assisted cancer diagnosis has been focused on diagnostic accuracy [7, 8, 9]. Deep learning methods such as Convolutional Neural Networks have been used to develop automated diagnostic systems to improve the traditional manual assessment of histopathological images by pathologists. Research has shown that a deep learning-based diagnostic system can help a human pathologist to improve diagnostic accuracy [8]. Such a diagnostic system has a great potential to support under-resourced healthcare facilities, which often have severe shortages of trained pathologists, to expedite diagnosis. Despite the technical advances, the real-world application in developing countries is hindered by a lack of computational resources and general inadequacy of healthcare infrastructures. This research focuses on designing a diagnostic system with computational efficiency and mobile readiness while achieving high diagnostic accuracy.

MobileNetV2, a lightweight CNN architecture, is designed to achieve computational efficiency using the Inverted Residual Structure and the Depthwise Separable Convolution [13]. The model size and complexity of MobileNetV2 is substantially lower than VGG16, ResNet50, and ResNet101 [16]. The high computational efficiency of MobileNetV2 makes it suitable for mobile devices or devices with low computational power. This research has demonstrated that the MobileNetV2-based diagnostic models outperformed its more complex counterparts VGG16, ResNet50 and ResNet101 in diagnostic performance and computational efficiency. The MobileNetv2 diagnostic models had a higher ROC AUC and a stronger model generalization capacity as compared with VGG16, ResNet50 and ResNet101. The MobileNetV2 models reduced the time per inference step by 66.8%, 59.5%, and 73.2% as compared with VGG16, ResNet50, and ResNet101, respectively. The results justify the application of MobileNetV2 in an IT resource-constrained environment. Developing countries are experiencing severe shortages of trained pathologists, which causes long delays of metastatic breast cancer and high mortality rates. The average number of pathologists per head of population is 1 to 1,000,000 in sub-Saharan regions, compared with the ratio of 1 pathologist to 15,000–20,000 in the US and UK [4]. Meanwhile, diagnosis quality may be impacted by the limitations of resources and training. Therefore, the MobileNetV2-based diagnostic system provides an AI-driven, resource-efficient, automated, and standardized solution to help relieve the severe shortages of trained pathologists and reduce the long diagnosis delays.

To tackle the public health crisis of long diagnosis delays in developing countries, a major challenge is inadequate accessibility to healthcare facilities. As an example, in sub-Saharan Africa, more than 170 million people are more than 2 hours from the nearest regional hospital, and about 40% of the population is more than 4 hours away from a national hospital (Figure 5B) [5]. While the accessibility to health care facilities is limited across sub-Saharan Africa, this is particularly challenging in Northern sub-Saharan countries, lacking access to all tiers of health care facilities, including basic health post (Tier 1), health center (Tier 2), regional hospital (Tier 3) and central hospital (Tier 4) (Figure 5B). The countries with high breast cancer mortality and those with severe health care resource shortage are geographically overlapped in Northern sub-Sahara (Figure 5A and B), suggesting a strong link between the elevated breast cancer mortality and limited health care resources. To address this challenge, the MobileNetV2-based diagnosis system can be integrated into the local healthcare network to facilitate histopathological image processing and assist pathologists' diagnosis. The system can be deployed to Tier 3 regional hospitals and Tier 4 central hospitals to assist pathology services (Figure 5C). Because the MobileNetV2 system is suitable for devices of low computational power, this system can also be deployed to Tier 2 and possibly Tier 1 healthcare facilities with limited IT resources. This system also connects lower tiers and high tiers healthcare facilities to expedite referrals for diagnosis of complex cases. This will shorten the delay in diagnosis, reduce the late-stage presentation at diagnosis, and therefore improve patients' survival outcomes.

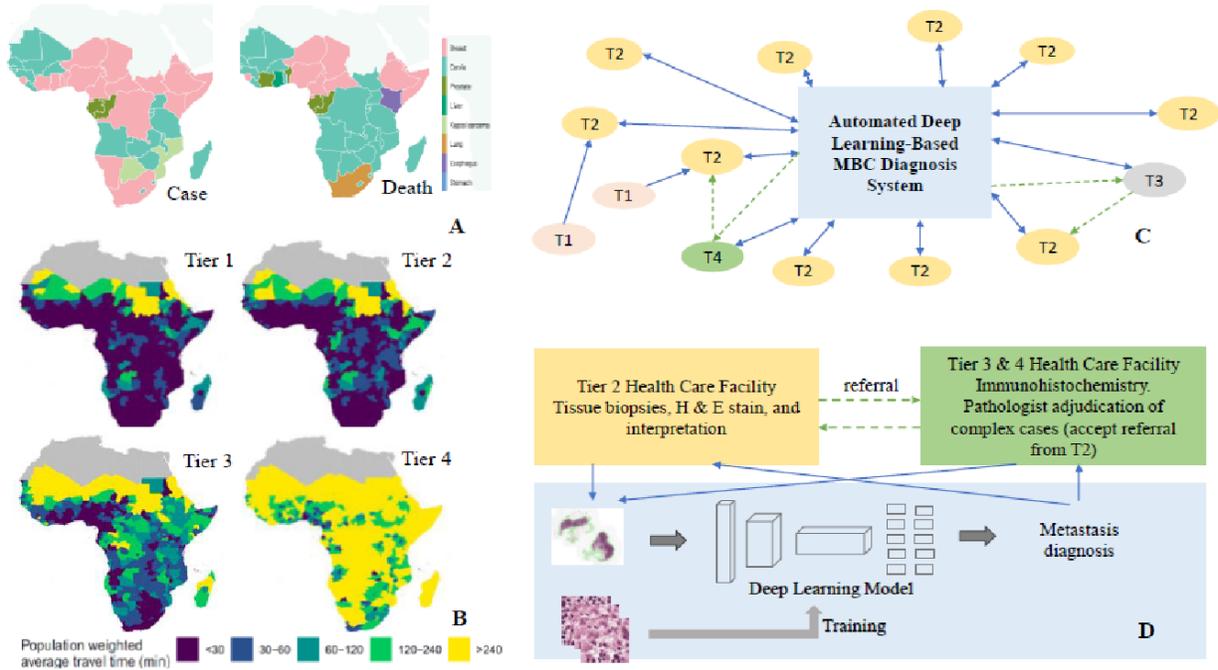

**Figure 5.** Prototypical schema for application of the deep learning-based automated diagnosis system in sub-Saharan Africa. (A) Sub-Saharan countries with high breast cancer cases (left) and high mortality rates (right) (both highlighted with pink). Breast cancer is the most common cancer in many sub-Saharan countries and has been the most common cause for cancer death in the Northern Sub-Saharan Africa [18].
(B) Accessibility to the nearest public health care facility (in travel hours), by Tier 1 to Tier 4 health care facility. Accessibility to public health care facilities is limited across the sub-Saharan region, especially in Northern sub-Saharan countries [5].
(C) The deep learning-based diagnosis system, highly efficient and mobile ready, can be deployed to all tiers of healthcare facilities. The collaborative diagnostic system can integrate knowledge from many institutions to support Tier 2 and possibly Tier 1 health care facilities for breast cancer diagnosis.
(D) The system also provides connectivity between different tiers of health facility to expedite diagnosis of complex cases.

## 5. Conclusion

To tackle the healthcare disparities in metastatic breast cancer diagnosis, this research has developed a deep learning-based diagnosis system for metastatic breast cancer that aims to achieve high diagnostic accuracy as well as computational efficiency suitable for an under-resourced environment. We evaluated four CNN architectures: MobileNetV2, VGG16, ResNet50 and ResNet101. The MobileNetV2-based diagnostic model outperformed the more complex VGG16, ResNet50 and ResNet101 models in diagnostic accuracy, model generalization, and model training efficiency. The visual comparison between the model prediction and ground truth has demonstrated that the MobileNetV2 diagnostic models can identify very small cancerous nodes embedded in a large area of normal cells which is challenging for manual image analysis. Equally Important, the light weighted MobleNetV2 models were computationally efficient and ready for mobile devices or devices of low computational power. These advances empower the development of a resource-efficient and high performing AI-based metastatic breast cancer diagnostic system that can adapt to under-resourced healthcare facilities in developing countries. This research provides an innovative technological solution to address the long delays in metastatic breast cancer diagnosis and the consequent disparity in patient survival outcome in developing countries.

Funding: This research received no specific grant from any funding agency in the public, commercial, or not-for-profit sectors.

Declaration Of Conflicting Interests:
The author(s) declared no potential conflicts of interest with respect to the research, authorship and/or publication of this article.

## 6. References

1. Sung, H., Ferlay, J., Siegel R.L., et al., 2021. Global cancer statistics 2020: GLOBOCAN estimates of incidence and mortality worldwide for 36 cancers in 185 countries. CA Cancer J. Clin. 71: 209-249.
2. Jedy-Agba, E., McCormack, V., Adebamowo, C., & Dos-Santos-Silva, I. 2016. Stage at diagnosis of breast cancer in sub-Saharan Africa: a systematic review and meta-analysis. The Lancet. Global health, 4(12), e923–e935. https://doi.org/10.1016/S2214-109X(16)30259-5.
3. Pace, L. and Shulman, L.N. 2016. Breast cancer in Sub-Saharan Africa: challenges and opportunities to reduce mortality. 21: 739-744.
4. Fleming, K. (2019) Pathology and cancer in Africa. ecancer, 13: 945. https://doi.org/10.3332/ecancer.2019.945
5. Falchetta, G., Hammad, A.T., and Shayegh, S. 2020. Planning universal accessibility to public health care in sub-Saharan Africa. PNAS. 117 (50): 31760 – 31769.
6. Rajpurkar, P., Chen, E., Banerjee, O., & Topol, E.J. (2022) AI in health and medicine. Nature Medicine. 28: 31-38.
7. Freeman, K., Geppert, J., Stinton, C., et al. (2021) Use of artificial intelligence for image analysis in breast cancer screening programs: systematic review of test accuracy. The BMJ. 374: n1872.
8. Wang, D., Khosla, A., Gargeya, R., et al. 2016. Deep learning for identifying metastatic breast cancer. arXiv: 1606.05718v1.
9. Veta, M., Heng, Y., Stathonikos, N., et al. 2019. Predicting breast tumor proliferation from whole-slide images: The TUPAC16 Challenge. Med. Image Anal. 54: 111-121.
10. Bejnordi, B.E., Litjens, G., Timofeeva, N, et al. 2016. Stain specific standardization of whole-slide histopathological images. IEEE transactions on medical imaging, vol. 35, no. 2, pp. 404–415.
11. Otsu, N. A threshold selection method from gray-level histograms. 1975. Automatica, vol. 11, no. 285-296, pp. 23–27.
12. Sural, S., Qian, G., and Pramanik, S. 2002. Segmentation and histogram generation using the HSV color space for image retrieval. in Image Processing. 2002. Proceedings. 2002 International Conference on, vol. 2. IEEE, 2002, pp. II–589.
13. Sandler, M., Howard, A., Zhu, M., et al. 2019. MobileNetV2: Inverted residuals and linear bottlenecks. arXiv: 1801.04381v4.

14. Simonyan, K. and Zisserman, A. 2015. Very deep convolutional networks for large-scale image recognition. arXiv: 1409.1556v6.
15. He, K., Zhang, X., Ren, S., et al. 2015. Deep residual learning for image recognition. arXiv: 1512.03385v1.
16. Keras API Reference (https://keras.io/api/applications/)
17. Shorten, C. and Khoshgoftaar, T.M. 2019. A survey on image data augmentation for deep learning. Journal of Big Data. 6: 60.
18. American Cancer Society. 2019. The Cancer Atlas.